From XDJ@mitlns.mit.eduTue Oct 15 13:43:24 1996
Date: Tue, 15 Oct 1996 13:41:15 -0400 (EDT)
From: XDJ@mitlns.mit.edu
To: xdj@quark.umd.edu

\documentstyle[prl,aps,preprint]{revtex}

\begin{document}
\tighten

\title{{HUNTING FOR THE REMAINING SPIN IN THE NUCLEON}
\thanks {Plenary talk given at the 12th International Symposium 
on High-Energy Spin Physics, Amsterdam, Sept. 1996. 
This work is supported in part by funds provided by the
U.S.  Department of Energy (D.O.E.) under cooperative agreement
DOE-FG02-93ER-40762.}}

\author{Xiangdong Ji}
\bigskip

\address{
Department of Physics \\
University of Maryland \\
College Park, Maryland 20742 \\
{~}}

\date{U. of MD PP\#97-042 ~~~DOE/ER/40762-102~~~ October 1996}

\maketitle

\begin{abstract}
This talk consists of four parts. In part one, I give an 
elementary discussion on constructing a Lorentz-invariant
spin sum rule for the nucleon. In part two, I discuss 
a gauge-dependent spin sum rule, explore its relation 
with the polarized gluon distribution, and introduce
the complete evolution equation for the spin structure. 
In part three, I consider a gauge-invariant spin sum rule
and the related evolution equation. The solution of the equation
motivates the possibility that half of the nucleon spin
may be carried by gluons at low energy scales. In the final
part, I discuss deeply-virtual Compton scattering as 
a possible way to measure the canonical orbital
angular momentum of quarks in the nucleon. 

\end{abstract}
\pacs{xxxxxx}

\narrowtext

Yesterday and today, we have heard essentially two kinds
of explanations to the so-called ``spin crisis''\cite{talks}. 
The first
kind says that the experimental data do not rule out 
the simple quark model prediction that 
the quark spin carries a large fraction of the nucleon spin. 
One way to see this is that the deep-inelastic sum rule has
an unknown uncertainty about the small $x$ contribution. 
Another way to see this is that one has to subtract 
the anomaly contribution from
the measured $\Delta \Sigma$ before comparing it with
the quark model prediction, and the subtraction is potentially 
large. The second kind of explanations is that the quark 
spin carries little of the nucleon spin, due to for instance 
a large negative sea polarizations. On the other hand, in the skyrme 
model discussed by J. Ellis, it seems that the majority 
of the nucleon spin is carried by orbital angular momentum. 
Whatever position one may take, it is safe to conclude 
that the nucleon spin carried by other sources is 
significant. Thus, in my talk, I will concentrate
on the subject of the remaining spin in the nucleon, i.e.
the part not measured by the polarized deep-inelastic
scattering experiment. 

\section{Construct a Lorentz-Invariant Spin sum rule}

To understand what are the remaining components of the nucleon
spin, it is important to construct a {\it Lorentz-invariant} 
spin sum rule. At first, it appears difficult to talk about
different contributions to the nucleon spin because in field theory
angular momentum operators do not commute with boost operators. 

States of a spin-1/2 particle are labelled by 4-momentum
$p^\mu$ and polarization vector $s^\mu$. Hence we write the
nucleon states as $|p,s\rangle$. To talk about spin in a 
relativistic way, one has to introduce the relativistic 
spin operator $\hat W_\mu$, which is also called the Pauli-Lubanski
spin, 
\begin{equation}
       \hat W_\mu \sim \epsilon_{\mu\alpha\beta\gamma}
    \hat J^{\alpha\beta} \hat P^\gamma \ , 
\end{equation}
where $\hat J^{\alpha\beta}$ are the generators of Lorentz transformations
and $\hat P^\mu$ is the energy-momentum operator. The fact that the
nucleon has spin 1/2 in all frames is represented by the following
equation, 
\begin{equation}
          \hat W^2 |ps\rangle = {1\over
2}\left({1\over2}+1\right)|ps\rangle \ . 
\end{equation}
Since $\hat W^2$ is quadratic in angular momentum and boost 
operators, the equation doesn't seem to offer 
any interesting spin sum rule.  

Notice, however, $s_\mu\hat W^\mu$ is also a Lorentz scalar and 
it has $|ps\rangle$ as its eigenstate,
\begin{equation}
       s_\mu\hat W^\mu |ps \rangle = {1\over 2}|ps\rangle\ . 
\end{equation}
Or, we can write, 
\begin{equation}
       {1\over 2} = \langle ps|s_\mu \hat W^\mu |ps\rangle\ , 
\end{equation}
where I have been casual about the normalization. The
equation can be used to construct spin sum rules:
If the Pauli-Lubanski
spin is a sum of several contributions,
$\hat W^\mu = \sum_i \hat W^\mu_i$, we can write,
\begin{equation}
      {1\over 2} = \sum_i ~\langle ps|s_\mu\hat W^\mu_i|ps\rangle\ . 
\label{sum}
\end{equation}
The above equation contains the boost operators in general. 
However, if one chooses $\vec{s}$ to be in the direction of $\vec{p}$,
which without loss of generality can be chosen to be the $z$ axis,
then,
\begin{equation}
           s_\mu \hat W^\mu \sim \hat J^{xy} \equiv \hat J^z\ , 
\end{equation}
where $\hat J^z$ is the $z$ component of the angular momentum 
operator. The nucleon is now in the helicity eigenstate $\lambda=1/2$, 
and a helicity sum rule emerges from Eq. (\ref{sum}), 
\begin{equation}
     {1\over 2} = \sum_i \left\langle p {1\over 2}
   \left|\hat J^z_i\right|p{1\over 2}\right\rangle \ , 
\end{equation}
where $\sum_i \hat J_i^z=\hat J^z$. This sum rule is most suitable
for studying the spin structure of the nucleon \cite{jaffemanohar}. 

To actually construct a spin sum rule, one needs to 
know the angular momentum operators in QCD, which are identified
as the generators of spatial rotations. By Noether's 
theorem, we can derive these from the transformation property
of the QCD lagrangian density under rotations.
Depending upon the final
form of the angular momentum operators one prefers to take, 
both gauge-dependent and gauge-invariant
sum rules can result.

\section{A gauge-dependent sum rule}

In a 1989 paper, Jaffe and Manohar wrote down the following
form of the QCD angular momentum operator \cite{jaffemanohar},
\begin{eqnarray}
      \vec{J} &=& \int d^3\vec{x} ~\Big[~
          {1\over 2}\bar \psi \vec{\gamma}\gamma_5\psi 
          + \psi^\dagger \vec{x}\times (-i\vec{\bigtriangledown})\psi
              \nonumber \\
          &+& \vec{E}\times \vec{A} 
          + E_i(\vec{x}\times \vec{\bigtriangledown})A_i ~\Big]\ .
\label{ang}
\end{eqnarray}
An advantage of this form is that the physical meaning of the individual
terms is quite obvious: The first term is the 
quark spin, the second term is the quark 
orbital angular momentum, the third term is the gluon spin, 
and the final term is the gluon
orbital angular momentum. According to the above equation, one 
can write down a sum rule for the nucleon spin,
\begin{equation}
        {1\over 2} = {1\over 2}\Delta \Sigma(\mu^2)
             + L_q'(\mu^2) + \Delta g(\mu^2) + L_g'(\mu^2) \ , 
\end{equation}
where, for instance,
\begin{equation}
     \Delta g(\mu^2) = \langle ps|\int d^3\vec{x} (\vec{E}\times\vec{A})^z|ps\rangle\ , 
\end{equation}
etc. Clearly, $L_q'$, $\Delta g$ and $L_g'$ are gauge, and hence frame, 
dependent.

Interestingly, $\Delta g$ in the infinite momentum frame
and light-like gauge ($A^+=0$) is related to a quantity 
present in polarized high-energy scattering,
\begin{equation}
     \Delta g(\mu^2) = \int^1_0 \Delta G(x,\mu^2) dx
\end{equation}
where $\Delta G(x,\mu^2)$ is the polarized gluon distribution. Recently, 
there has been a lot of discussion in the literature 
about measuring $\Delta G(x)$ at polarized RHIC and HERA. 
I am happy to see that there will be a round table discussion
about this topic on Friday. 

The individual contributions to the nucleon spin are scale-dependent.
Recently, Hoodbhoy, Tang and myself \cite{ji1} have worked out the 
scale dependence of the orbital angular momentum contributions. 
This subject was first recognized by Phil Ratcliffe \cite{rat}. 
Together with the well-known Altarelli-Parisi equation \cite{ap}, 
we now have a complete set of equations to evolve 
the spin structure of the nucleon at the leading-log level, 
\begin{equation}
{\partial \over \partial \ln \mu^2}
  \left(\begin{array}{c}
        \Delta \Sigma(\mu^2) \\
        \Delta g(\mu^2) \\
         L_q'(\mu^2) \\
          L_g'(\mu^2) 
    \end{array} \right) 
   = {\alpha_s(\mu^2)\over 2\pi}
    \left( \begin{array}{rrrr}
         0 & 0 &0 &0 \\
         {3\over 2}C_F &{\beta_0\over 2} &0 &0 \\
        -{2\over 3}C_F &{n_F\over 3} &-{4\over 3}C_F &{n_F \over 3} \\
        -{5\over 6}C_F& -{11\over 2}& {4\over 3}C_F &-{n_F \over 3} \\
      \end{array} \right) 
       \left( \begin{array}{c}
             \Delta \Sigma(\mu^2) \\
           \Delta g(\mu^2) \\
           L_q'(\mu^2) \\
            L_g(\mu^2) 
        \end{array} \right) \ .  
\end{equation}
If one knows the decomposition of the spin of the nucleon
at one perturbative scale, one can solve from the above equation 
the decomposition at any other perturbative scale. 
As $\mu^2\rightarrow \infty$, one has the following 
asymptotic solution, 
\begin{eqnarray}
      &&\Delta \Sigma \rightarrow {\rm const.} \nonumber \\
      &&\Delta g \rightarrow {\lambda}\ln \mu^2 + {\rm const.} \nonumber
      \\ 
      && L_q' \rightarrow {\rm const.} \nonumber \\
      && L_g' \rightarrow {\rm -\lambda \ln \mu^2 + const.}
\end{eqnarray}
Thus, the gluon helicity increases logarithmically with
the probing scale. That increase is entirely cancelled
by the gluon orbital contribution in the asymptotic 
limit. 

\section{ a gauge-invariant sum rule}
Recently, I have proposed to reorganize the angular momentum 
operator in Eq. (\ref{ang}) so that it is explicitly
gauge-invariant \cite{ji2},  
\begin{eqnarray}
     \vec{J} &=& \int d^3 \vec{x}~
        \Big[~ {1\over 2}\bar \psi \vec{\gamma}\gamma_5 \psi \nonumber \\
          & + & \psi^\dagger (\vec{x}\times (-i\vec{D}))\psi \nonumber \\
          & + &\vec{x}\times(\vec{E}\times\vec{B}) ~\Big] \ .
\end{eqnarray}
As before, the first term is the quark spin. The
second term, in which the covariant derivative is  
${\vec D} = \vec{\partial} + ig\vec{A}$, is the
canonical orbital angular momentum of quarks. 
The last term is the angular momentum of the gluons, as 
is clear from the appearance of the Poynting vector. 
According to the above, we can
write down a gauge-invariant spin sum rule, 
\begin{equation}
      {1\over 2} = {1\over 2}
     \Delta \Sigma(\mu^2) + L_q(\mu^2) + J_g(\mu^2) \ , 
\end{equation}
where the second and third terms are quark orbital and
gluon contributions, respectively. I introduce the sum of 
the first and second terms as $J_q(\mu^2)$, representing the total
quark contribution. It is interesting to notice that 
although $\Delta \Sigma(\mu^2)$ is affected by the
axial anomaly, $J_q(\mu^2)$ is anomaly-free \cite{ji1}. 
 
The evolution equation 
for the quark and gluon contributions is, 
\begin{equation}
{\partial \over \partial \ln \mu^2}
  \left(\begin{array}{c}
         J_q(\mu^2) \\
          J_g(\mu^2) 
    \end{array} \right) 
   = {\alpha_s(\mu^2)\over 2\pi}
    {1\over 9}\left( \begin{array}{rr}
        -16 & 3n_F  \\
        16 & -3n_F  \\
      \end{array} \right) 
        \left( \begin{array}{c}
           J_q(\mu^2) \\
            J_g(\mu^2) 
        \end{array} \right) \ . 
\end{equation}
As $\mu^2\rightarrow \infty$, there is a fixed point
solution, 
\begin{eqnarray}
       J_q(\infty) &=& {1\over 2} {3n_f\over 16 + 3n_f} \ ,  \nonumber \\
       J_g(\infty) &=& {1\over 2} {16\over 16 + 3n_f} \ . 
\end{eqnarray}
Thus we see about half of the nucleon spin is carried by
gluons. A similar result was obtained by Gross and
Wilczek in 1974 for the quark and gluon contributions to 
the momentum of the nucleon \cite{gw}. Experimentally, 
one finds that about half of the nucleon momentum is 
carried by gluons already at quite low-energy scales. 
An interesting question is whether the gluons carry half
of the nucleon spin at low energy scales?  

It is difficult to answer this question theoretically, 
because QCD is difficult to solve. Recently, Balitsky
and I made an estimate using the QCD sum rule approach
\cite{jibalitsky}. 
We find, 
\begin{equation}
    J_g(\mu^2\sim 1 {\rm GeV}^2) \simeq {4\over 9} {e<\bar u\sigma Gu>
     <\bar uu> \over M_{1^{-+}}^2\lambda_N^2}
\end{equation}
which gives approximately
0.25. If this calculation indicates
anything about the truth, the spin structure of the nucleon
roughly looks like this,
\begin{equation}
       {1\over 2} = 0.10({\rm from~} {1\over2}\Delta \Sigma) 
        + 0.15({\rm from~} L_q) + 0.25({\rm from~} J_g) \ . 
\end{equation}
It would be interesting to test this scenario. 

\section{How to measure $J_{q,g}$?}
By examining carefully the definition of the
matrix elements,
\begin{equation}
      J_{q,g}(\mu^2) = \langle p{1\over 2} \left|
         \int d^3x (\vec{x}\times \vec{T}_{q,g})^z
 \right|p{1\over 2}\rangle \ , 
\label{matrix}
\end{equation}
one realizes that they can be extracted from the 
form factors of the quark and gluon parts of 
the QCD energy-momentum tensor $T^{\mu\nu}_{q,g}$. 
Using Lorentz symmetry, we can write down the 
forward matrix elements of $T^{\mu\nu}_{q,g}$, 
\begin{eqnarray}
      \langle p'| T_{q,g}^{\mu\nu} |p\rangle 
       &=& \bar u(p') \Big[A_{q,g}(\Delta^2)  
       \gamma^{(\mu} \bar P^{\nu)} + 
   B_{q,g}(\Delta^2) \bar P^{(\mu} i\sigma^{\nu)\alpha}\Delta_\alpha/2M \nonumber \\
   &&  +  C_{q,g}(\Delta^2)(\Delta^\mu \Delta^\nu - g^{\mu\nu}\Delta^2)/M   
   + \bar C_{q,g}(\Delta^2) g^{\mu\nu}M\Big] u(p)\ , 
\end{eqnarray}
where $\bar p^\mu=(p^\mu+{p^\mu}')/2$, $\Delta^\mu 
= {p^\mu}'-p^\mu$, 
and $u(p)$ is the nucleon spinor. 
Taking the forward limit in the $\mu=0$ component and integrating
over 3-space, one finds that $A_{q,g}(0)$ give 
the momentum fractions of the nucleon carried by 
quarks and gluons ($A_q(0)+A_g(0)= 1$). 
On the other hand, substituting 
the above into the nucleon matrix element of Eq. (\ref{matrix}),  
one finds \cite{ji2},  
\begin{eqnarray}
      J_{q, g} = {1\over 2} \left[A_{q,g}(0) + B_{q,g}(0)\right] \ . 
\end{eqnarray}
There is an analogy for this. If one knows the Dirac and Pauli
form factors of the electromagnetic current, $F_1(Q^2)$ and $F_2(Q^2)$, 
the magnetic moment of the nucleon, which is defined as 
the matrix element of (1/2)$\int d^3x (\vec{x} \times \vec{j})^z$ , 
is just $F_1(0) +F_2(0)$. 
 
How to measure the form factors of the energy momentum tensor? 
If one has two vector currents which are separated along 
the light-cone, it is known from the operator product 
expansion that, 
\begin{equation}
       TJ_\alpha(z)J_\beta(0) \rightarrow ... +
C_{\alpha\beta\mu\nu}(z^2) T^{\mu\nu} + ....
\end{equation}
Thus to get the matrix element $\langle p'|T^{\mu\nu}|p\rangle$, 
we need $\langle p'|TJ_\alpha(z)J_\beta(0)|p\rangle$, i.e.  
a Compton scattering amplitude. 
To ensure the separation of the two currents is along the light-cone,
we let one of the photon momenta approach the Bjorken limit. 
Then it is easy to 
show that the Compton scattering is dominated by the single quark 
process. I shall call such a scattering process 
deeply-virtual Compton scattering (DVCS). 
 
What does one learn from DVCS? An analysis shows that one learns 
about the off forward parton distributions (OFPDs), 
which are defined through the following light-cone correlations,
\begin{eqnarray}
 \int  {d\lambda \over 2\pi} e^{i\lambda x}
      \langle p'|\bar\psi(-{\lambda n/ 2})\gamma^\mu
            \psi(\lambda n/2)|p \rangle
        &=& H(x,\Delta^2, \xi) \bar u(p')\gamma^\mu u(p) \nonumber \\
         && + E(x,\Delta^2, \xi) \bar u(p'){i\sigma^{\mu\nu}
             \Delta_{\nu}
          \over 2M}u(p) + ...  \nonumber \\
 \int  {d\lambda \over 2\pi} e^{i\lambda x}
      \langle p'|\bar\psi(-{\lambda n/ 2})\gamma^\mu\gamma_5
            \psi(\lambda n/2)|p \rangle
      & =& \tilde H(x,\Delta^2, \xi) 
        \bar u(p')\gamma^\mu \gamma_5 u(p) \nonumber \\
       && + \tilde E(x, \Delta^2, \xi) \bar u(p')
          {\gamma_5\Delta^\mu
         \over 2M}u(p)
        + ...
\end{eqnarray}
where I have neglected the gauge link and the 
dots denote higher-twist distributions.
>From the definition, $H$ and $\tilde H$ are nucleon
helicity-conserving amplitudes and $E$ and $\tilde E$
are helicity-flipping. Such distributions 
have been considered in 
the literature before \cite{other}. 

The off-forward parton distributions have
the characters of both ordinary parton distributions and
nucleon form factors. In fact in the limit of
$\Delta^\mu \rightarrow 0$, we have 
\begin{equation}
     H(x,0,0) = q(x)\ ,~~~ \tilde H(x,0,0) = \Delta q (x) \ , 
\end{equation}
where $q(x)$ and $\Delta q(x)$ are quark and quark helicity
distributions. On the other hand, forming 
the first moment of the new distributions, one gets
the following sum rules \cite{ji2,other}, 
\begin{eqnarray}
     \int^1_{-1} dx H(x,\Delta^2, \xi) &=& F_1(\Delta^2) \ ,  \nonumber \\
     \int^1_{-1} dx E(x,\Delta^2, \xi) &=& F_2(\Delta^2) \ . 
\end{eqnarray}
where $F_1$ and $F_2$ are the Dirac and Pauli form factors. 
The most interesting sum rule relevant to the nucleon spin is,
\begin{eqnarray}
     \int^1_{-1} dx x [H(x, \Delta^2, \xi) +
       E(x, \Delta^2, \xi) ]
     = A_q(\Delta^2) + B_q(\Delta^2) \ ,  
\end{eqnarray} 
where luckily the $\xi$ dependence, or $C_q(\Delta^2)$
contamination, drops out. Extrapolating the sum rule 
to $\Delta^2=0$, the total quark (and hence quark orbital) 
contribution to the nucleon spin is obtained. By forming still 
higher moments, one gets form factors of various 
high-spin operators. 

There are a lot of theoretical and experimental questions about 
DVCS. Theoretical questions include: 
is there a factorization
theorem for DVCS? is there an Altarelli-Parisi equation for the OFPDs
evolution? what is the small $x$ and $\xi$ behavior? how to extrapolate
the form factors to $\Delta^2=0$? Experiment-related questions 
include: how big is the cross section? will the Bethe-Heitler process
overshadow DVCS? what kinematic region corresponds to DVCS?
does one need polarizations of beam? target? 
how practical is to form sum rules? etc. Some of these 
questions have been  
answered in recent papers \cite{ji3,ra}. Others are open.   

\acknowledgements
I thank the organizers of this meeting for the opportunity 
to discuss this interesting subject and for the remarkable effort
to arrange a visa in two days, which even surprised the 
Dutch Embassy in Washington DC! I thank Wally Melnitchouk
for a careful reading of this write-up.

\end{document}